\documentclass[aps,pre,twocolumn,showpacs]{revtex4-1}
\usepackage{amsmath,amsfonts,bm, color,amssymb}
\usepackage{graphicx,epsfig,overpic,hyperref,ulem}

\newcommand{\refeq}[1]{Eq. \ref{#1}}

\newcommand{\sigm}{{\sigma}}
\newcommand{\eps}{{\varepsilon}}
\newcommand{\out}{{\mathrm{s}}}
\newcommand{\ins}{{\mathrm{p}}}

\newcommand{\bE}{{\bf E}}
\newcommand{\bP}{{\bf{P}}}

\newcommand{\bOmega}{{\bm{\Omega}}}
\newcommand{\Rr}{\mbox{\it R}}
\newcommand{\Sr}{\mbox{\it S}}
\newcommand{\mw}{{\mathrm{mw}}}

\begin{document}
\title{Tuning the random walk of active colloids} 

\author{Hamid Karani\textit{$^{a}$}, Gerardo E.~Pradillo\textit{$^{b}$}, Petia M.~Vlahovska\textit{$^{a,b}$}}
\affiliation{%
$^{a}$~Engineering Sciences and Applied Mathematics, Northwestern University, Evanston, IL 60208, USA. E-mail: petia.vlahovska@northwestern.edu\\
$^{b}$~Mechanical Engineering, Northwestern University, Evanston, IL 60208, USA\\ }

\date{\today}

\begin{abstract}
Active particles such as swimming bacteria or self-propelled colloids are known to spontaneously organize into fascinating large-scale dynamic structures. The emergence of these collective states from the motility pattern of the individual particles, typically a random walk, is yet to be probed in a well-defined synthetic system. 
Here, we report the experimental realization of  intermittent colloidal motion that reproduces the run-and-tumble and L\'{e}vy trajectories
 common to many swimming and swarming bacteria.   
Our strategy enables to tailor the sequence of  repeated ``runs" (nearly constant-speed straight-line translation) and  ``tumbles" (seemingly erratic turn) to emulate any random walk.
 This new paradigm for active locomotion at the microscale opens new opportunities for experimental explorations 
 of the collective dynamics emerging in active suspensions. We find that population of these  random walkers exhibit behaviors reminiscent of bacterial suspensions such as dynamic clusters and mesoscale turbulent-like flows. 

\end{abstract}

\maketitle

Swimming bacteria navigate their environment by executing random walks \cite{Elgeti15, Lauga:2016}. 
 In a general run-and-turn type,  persistent swimming is interrupted by random changes in the direction of motion.
Examples include the widely studied run-and-tumble  {\it{E. coli}} \cite{Berg72}, run-and-active-stop  {\it{R. sphaeroides}} \cite{Pilizota09}, run-and-reverse {\it{P. putida}} \cite{Theves13}, and run-reverse-flick {\it{V. alginolyticus}} \cite{Xie11}. 
These motility strategies have inspired great interest in the engineering 
of artificial self-propelled particles that mimic the elaborate 
locomotion patterns of their biological counterparts \cite{Ebbens:2018, Han:2018, Palagi:2018, Huang:2019}. 
Most available experimental designs of artificial colloidal microswimmers perform active Brownian motion \cite{Paxton04,Bidoz05,Howse07,Jiang10,Buttinoni12,Baraban13,Samin15,Gomez16,Narinder18}, where the reorientation in the directed motion is driven by the rotational diffusion of the swimmer. This results in slow and continuous directional changes, in contrast to the sudden turning events characteristic of the run-and-tumble bacteria. 
Efforts to emulate  the kinetics of the bacterial run-and-tumble 
 motions \cite{Ebbens10,Ebbens12,Mano17}  have been unable to achieve truly random reorientation events with controllable turn time. 
Only recently, reorientation disentangled from rotational diffusion has been accomplished by triggering elastic recoil in a non-Newtonian fluid suspending the motile colloid \cite{Lozano18}. However, this approach does not allow to change the duration of the turn step as it is set by the elastic relaxation  time of the fluid.

\begin{figure*}[t]
	\includegraphics[width=\textwidth]{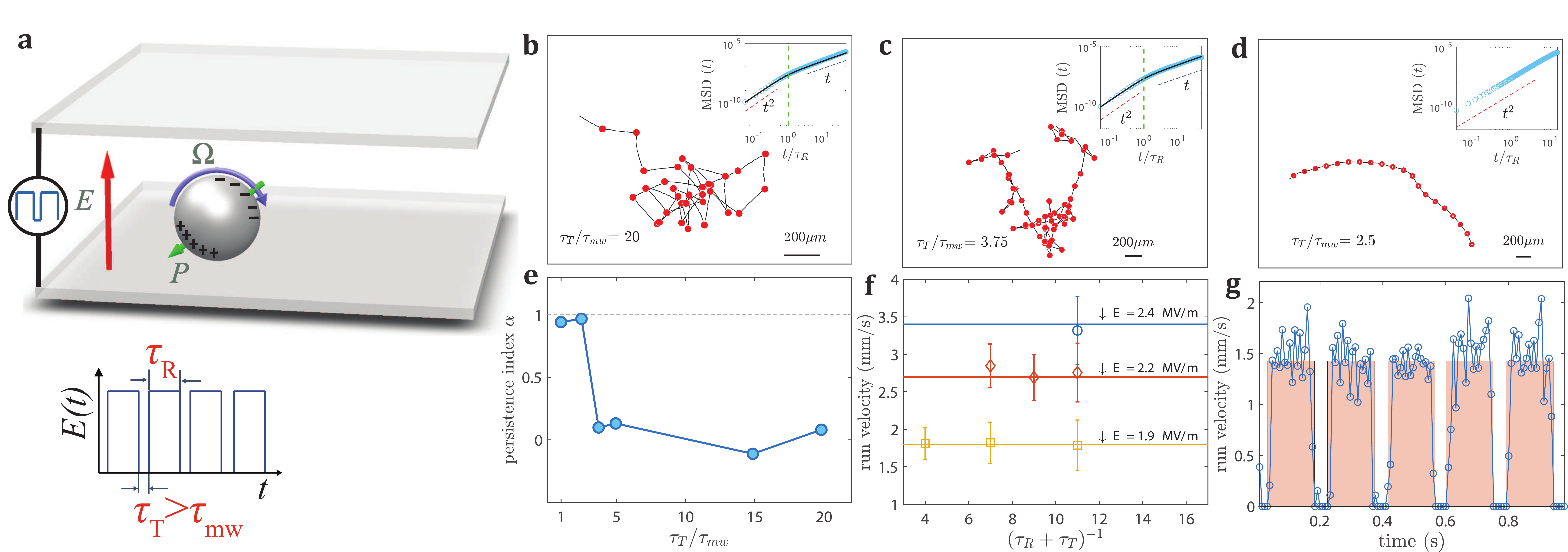}
	\caption{ \footnotesize {
(a) Quincke rotation: In a uniform dc electric field, free charges brought by conduction accumulate at the particle surface.
A spontaneous symmetry breaking  of the charge distribution  gives rise to a
net torque and the sphere spins at a rate $\bOmega$ about a randomly chosen axis in the plane perpendicular to the applied field direction. Experimental set-up: An insulating colloidal sphere is resting on the bottom electrode of a parallel-plate chamber filled with  weakly-conducting oil. A uniform electric field $E$ with magnitude above the threshold for Quincke rotation is applied  to cause the sphere to roll. If a square-wave electric field is applied with period between the pulses longer  than the time needed for the sphere to depolarize (Maxwell-Wagner time), the sphere executes a random walk.
(b-d) Quincke walker trajectories at different $\tau_T/\tau_\mw$ ratios, with $\tau_R=0.15$ s. Particle stops are marked with red circles. Insets: log-log plot of time-averaged experimental (symbol) and theoretical (solid line) mean-square displacement (crossover marked by the vertical dashed line at $t=\tau_R$). (e) persistence index $\alpha$ for different $\tau_T/\tau_\mw$ ratios.
(f) Run velocity  depends solely on the amplitude of the applied electric field. Symbols: measured velocity at different amplitude and frequencies  $ 1/(\tau_R + \tau_T)$ of the applied pulsed signal; Solid lines: velocity measured in dc fields. (g) particle velocity for 1 second duration of the pulsed signal, $\tau_T/\tau_\mw=20$, $\tau_R=0.15$ s.  E=1.66 MV/m.
}}
	\label{fig1}
\end{figure*}

\begin{figure*}[t]
	\includegraphics[width=\textwidth]{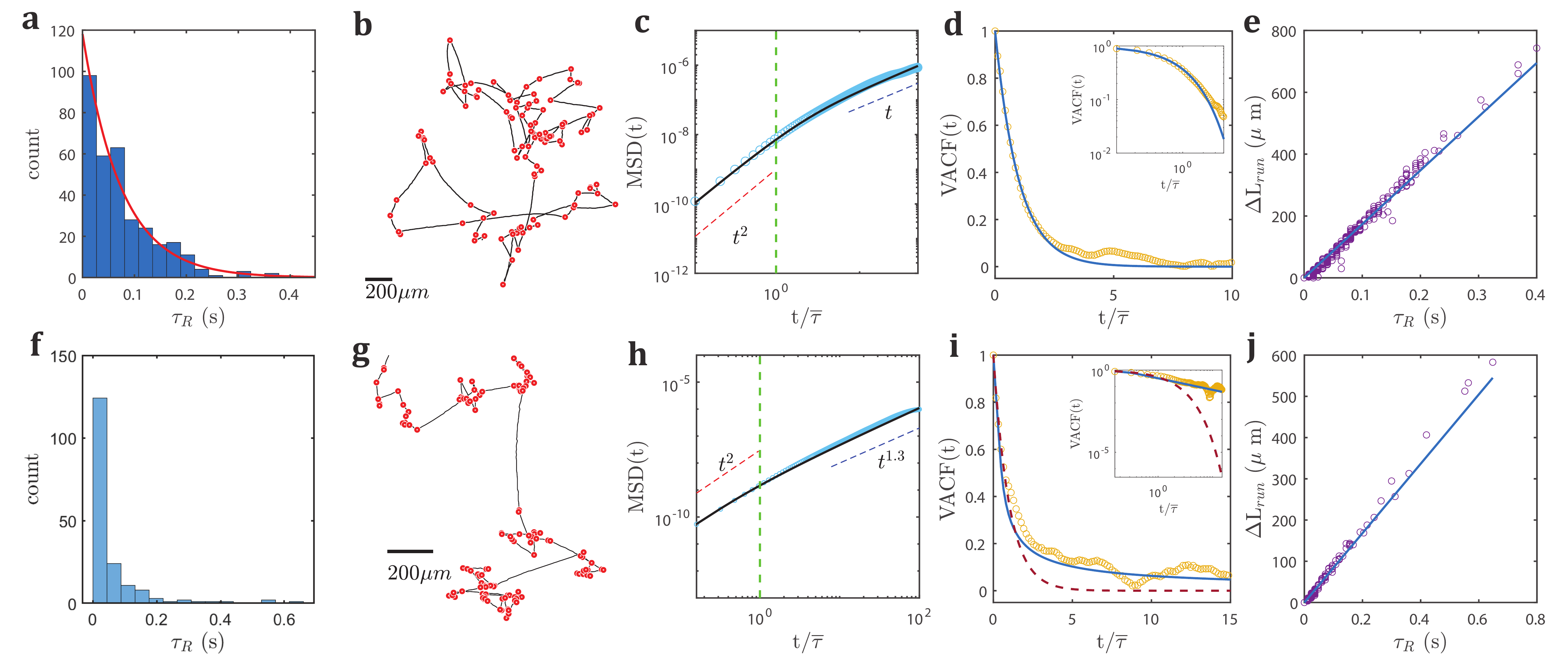}
		\caption{\footnotesize{Top row: Run-and-tumble, Bottom row: L\'{e}vy walk; (a,f) distribution of run-time $\tau_R$ drawn from an exponential and power-law PDF. (b,g) trajectory of the Quincke walker performing a run-and-tumble and L\'{e}vy walk. (c,h) time-averaged mean-square displacement. Symbols: experiment, Solid line: theory, Vertical dashed line marks $\overline{\tau}$. (d,i) normalized velocity autocorrelation function, Symbols: experiment, Solid line: theory, dashed line -exponential decay. Inset: same in log-log scale plot. (e,j) Run length {\it vs.} the corresponding run time $\tau_R$. Symbols: experiment, Solid line: linear fit showing a constant run velocity of $V=1.73$ mm/s for Run-and-Tumble and $V=0.84$ m/s for L\'{e}vy walk. E$=1.83$ MV/m for run-and-tumble and E$=1.5$ MV/m for L\'{e}vy walks. For both run-and-tumble and L\'{e}vy walk: $\overline{\tau}=0.075$ s, $\tau_T/\tau_{\mw}=20$. $\gamma=1.7$ for L\'{e}vy walk.}}
	\label{fig2}
\end{figure*}

Here we report the experimental realization of a motile colloid, inspired by the Quincke roller \cite{Bricard13,Bricard15, Gerardo:2019}, that performs  finely--tunable, diverse random walks such as
run-and-tumble or L\'{e}vy  walks (Fig. \ref{fig1}b, Fig. \ref{fig2}b,g).
A population of the Quincke random walkers display collective dynamics reminiscent of bacterial suspensions such as self-organization into large--scale aggregates and turbulent--like flows. They form swarms, rotating clusters,  polar clusters cruising over the whole domain without significant exchange of particles and  dynamic disordered clusters that continuously  deform and break by exchanging particles.

The colloid ``run'' is powered by  Quincke rotation, i.e.,
the spontaneous spinning of a particle polarized in a uniform DC electric field
 \cite{Quincke:1896} (see Fig. \ref{fig1}a and Appendix for a detailed description of the phenomenon). If the sphere is on a surface, it rolls steadily following a straight trajectory. The Quincke rollers have stirred a lot of interest since they were
discovered to undergo collective directed  motion  \cite{Bricard13, Bricard15, Morin17, Geyer:2018}.  Our strategy to introduce a ``tumble'' in the colloid trajectory exploits a unique feature of the Quincke instability: the degeneracy of the rotation axis in the plane perpendicular to the applied electric field (and parallel to the rolling surface). A sequence of  on-off-on electric field causes the sphere to roll-stop-turn; the turn is due to the  Quincke instability picking a new axis of rotation. One caveat, though, is that
the charging and discharging of the particle occurs by conduction and require finite time. The induced dipole $\bP$ evolves as \cite{Lemaire:2002}
\begin{equation}
\frac{\partial \bP}{\partial t}=\bOmega \times \bP-\tau_\mw^{-1} \left(\bP-\chi_e \bE\right)\,.
\end{equation}
where $\bOmega$ is the rotation rate and $\chi_e$ is the electric susceptibility of the particle.
The characteristic time scale for polarization relaxation  is the Maxwell-Wagner time  $\tau_\mw=(\eps_\ins+2\eps_\out)/(\sigma_\ins+2\sigma_\out)$, which depends solely on the fluid and particle conductivities and permittivities, $\sigma$ and $\eps$.
Random reorientation after each run is only ensured if the sphere is completely discharged  before the field is turned on. Incomplete depolarization acts as a memory and correlates subsequent runs. Thus the relaxation nature of the polarization  adds  another functionality to the Quincke walks: variable degree of run correlation. Furthermore,
since the Maxwell-Wagner ``memory" time scale depends solely on the fluid and particle electric properties it can be tuned by  adding surfactants to the oil \cite{Gerardo:2019} in the range between milliseconds to seconds;  in our experimental system we set it to few milliseconds.

As proof-of-concept experiments, we apply external electric field by designing a sequence of electric pulses with duration $\tau_R$ and spaced in time by $\tau_T$  to dielectric (polysterene)  micron-sized spheres (diameter 40 $\mu$m) settled onto the bottom electrode of an oil -filled rectangular chamber (Fig. \ref{fig1}a) \cite{Gerardo:2019}. 
As predicted, various trajectories are realized depending on the degree of depolarization, i.e. $ \tau_T/\tau_\mw$ (Fig. \ref{fig1}b-d). 
If $\tau_T\gg \tau_\mw$,  particle polarization relaxes completely and full randomization of the consecutive run directions is accomplished. Run and turn phases are independent and the particle undergoes an unbiased and uncorrelated random walk (Fig. \ref{fig1}a). 
The time-averaged mean-square displacements show excellent quantitative agreement with the theoretical predictions (summarized in the Appendix). 
The transition from a ballistic   to diffusive motion occurs at time $t\sim\tau_R$ and the long-time  behavior follows
\begin{equation}\label{eq:msd_unf}
MSD(t) = {V^2} \tau_R^2 t /{(\tau_R+\tau_T)}\,, \quad t\gg \tau_R\,.
\end{equation}
Typical run velocities $V\sim$1 mm/s result in an effective diffusion coefficient  on the order of few mm$^2$/s, quite large for a microswimmer.
As $\tau_T$ approaches $\tau_\mw$, the colloid motion starts to exhibit some local directional bias (Fig. \ref{fig1}c). Eventually the random walk vanishes completely and the particle undergoes a persistent directed motion (Fig. \ref{fig1}d). The trajectory is curved instead of a straight line because particle density is nonuniform (due to presence of microbubbles).
The sharp transition from the uncorrelated random walk to directed motion is illustrated in Fig. \ref{fig1}e by the persistence index $\alpha = \langle\cos(\Delta\theta)\rangle$, which quantifies the average change in the direction of motion after a run.
$\Delta\theta$ is the angle between two consecutive run segments and $\langle.\rangle$ is the average over all reorientation events. 
The sharp transition around $\tau_T/\tau_\mw\sim 2$ highlights the fact that complete depolarization and re-polarization, each occurring on time scale $\sim\tau_\mw$, are necessary for randomization of direction of motion.  
Thus in our design for a random walker, any resting time $\tau_T$ sufficiently larger than $\tau_\mw$ guarantees full randomization; hence rendering it a suitable swimmer for versatile application with different locomotion time-scales.

The average run velocity is independent of the frequency of electric field signal and is equal to the velocity with which the particle cruises at time-independent dc field with the same magnitude(Fig. \ref{fig1}f). Therefore, run velocity can be controlled by the amplitude of the applied signal.
Closer inspection of the particle motion shows that the particle follows the applied electric signal during the run and rest phases  (Fig. \ref{fig1}g). 

{\it{Run-and-Tumble and L\'{e}vy walks:}}
We now proceed to construct more complex locomotion patterns such as run-and-tumble and L\'{e}vy walk.
Unlike the simple random walk, which is characterized by a constant run time, run times for run-and-tumble  motion  are exponentially, and  for  L\'{e}vy walks -- power-law, distributed. We  randomly draw run times $\tau_R$ from the corresponding distribution  $\psi_R$ (Fig. \ref{fig2}(a,f)) and encode this times as  pulse durations in the electric signal (sample signals are shown in Fig. \ref{fig:signal}). In run-and-tumble, $\psi_R = 1/\bar\tau e^{-t/\bar{\tau}}$, where $\bar \tau$ is the  mean value for the run times \cite{Berg08,Angelani13}. For L\'{e}vy walk with resting periods \cite{Zaburdaev15},  $\psi_R = \gamma t_0^\gamma t^{-(1+\gamma)} H(t-t_0)$, where $\bar\tau=t_0\gamma/(\gamma-1)$ \cite{Sato99,Angelani13}. 
$H$ is the Heaviside function and $t_0$ is the lower cutoff value for run times.
The power $1<\gamma<2$ controls the degree of anomalous super-diffusion manifested at long times. 
In both cases, we assume turning time $\tau_T$ and run velocity $V$ to be constant.

Sample trajectories of the Quincke roller performing a run-and-tumble motion and L\'{e}vy walk are shown in Fig. \ref{fig2}b,g. 
The measured mean squared displacement ( Fig. \ref{fig2}c,h)  displays a transition from the initial ballistic regime for times shorter than $\overline{\tau}$ to  final normal diffusion with a linear scaling with time in the case of run-and-tumble. The experimental results are in excellent agreement with the theoretical prediction \cite{Angelani13}:
\begin{equation}\label{eq:msd_exp}
MSD(t) = {2 V^2 \bar\tau^2} \left( e^{-t/\bar \tau} + t/\bar \tau -1   \right)/{\left(1+\tau_T/\bar \tau\right)}\,.
\end{equation}
The long-time MSD for L\'{e}vy walk exhibits superdiffusion with a power consistent with the theoretical scaling of $t^{3-\gamma}$.
The experimentally observed superdiffusion persists up to $t/\overline{\tau}\approx 100$, beyond which, it starts to deviate from the asymptotic theoretical scaling due to the under-sampling of longer stretches which are responsible for the anomalous superdiffusive behavior. 
Contrary to previous cases, enhancing the long time statistics by repeating the experiments over several realizations is not trivial as the chance of losing the particle from the field of view during one of the long excursions is very high. 
In order to further analyze the performance of the Quincke random walker, we measure the experimental velocity auto-correlation function VACF from the trajectory analysis. Fig. \ref{fig2}d shows a sharp decay of the VACF  in the case of run and tumble motion, in agreement with  the theoretical predictions: 
\begin{equation}\label{eq:vac_exp}
VACF(t) = {V^2} e^{-t/\bar\tau}/\left({1+\tau_T/\bar\tau}\right)\,.
\end{equation}
For L\'{e}vy walk, VACF exhibits a tail (Fig. \ref{fig2}i) which agrees well with the theoretical prediction for the L\'{e}vy walk  and shows a poor fit to an exponential curve which drops sharply to zero. This, plus the fact that particle's displacement follows the desired distribution, further corroborates that the walker undergoes a L\'{e}vy walk. 
The run length (Fig. \ref{fig2}e,j) shows linear dependence  on the corresponding run times, which confirms that the walker runs at almost constant speed.

\begin{figure*}[t]
	\includegraphics[width=\textwidth]{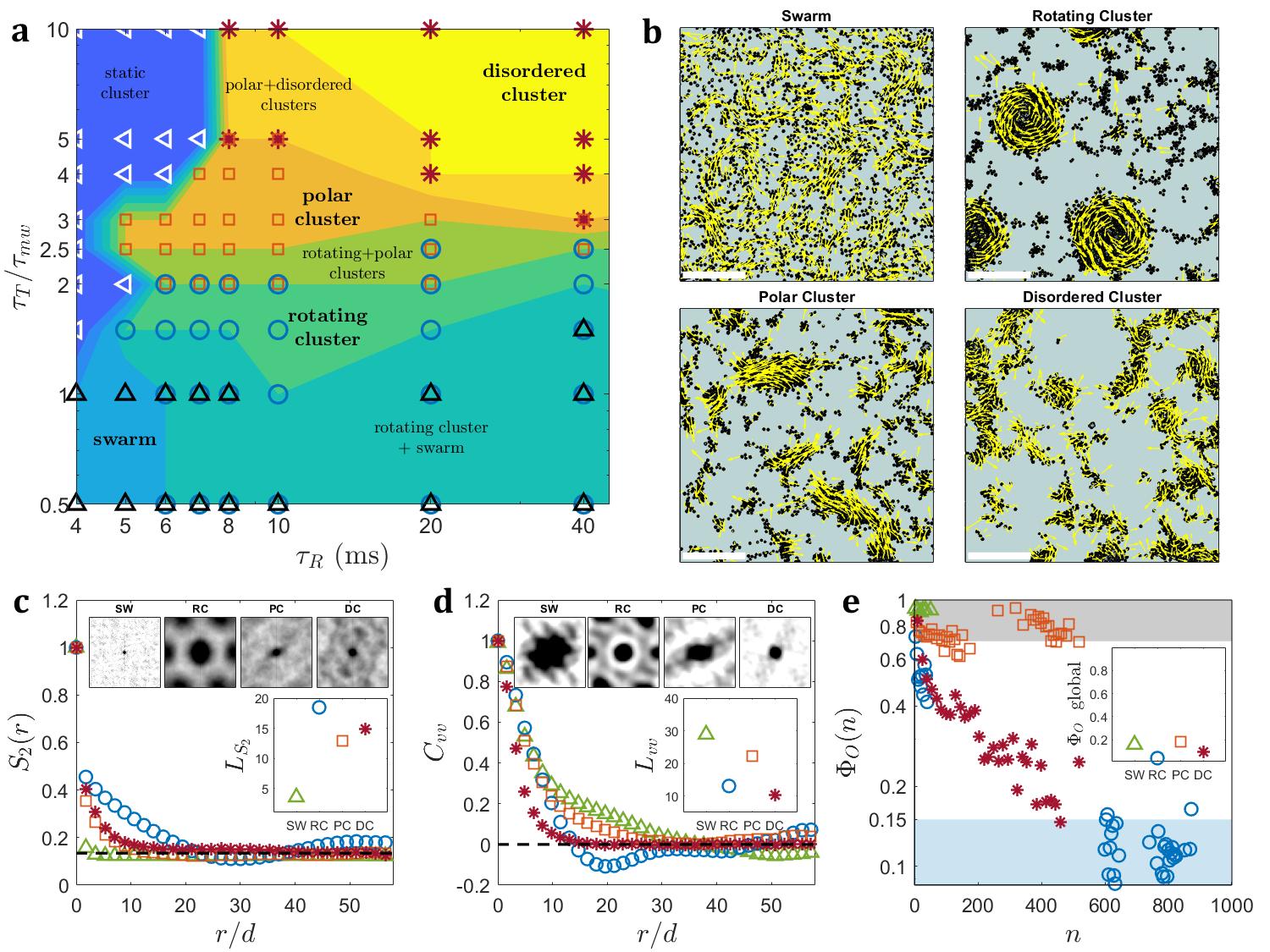}
		\caption{ \footnotesize{(a) Collective states formed by Quincke random walkers with different run and turn times ($\tau_R$, $\tau_T$). Symbols indicate the experimentally observed  $\lhd$: static cluster, $\bigtriangleup$: swarm (SW), $\circ$: rotating cluster (RC), $\square$: polar cluster (PC), and $\ast$: disordered cluster (DC).  		(b) Snapshots of  the SW ($\tau_T/\tau_\mw=0.5$,$\tau_R=4$ ms), RC ($\tau_T/\tau_\mw=1.5$,$\tau_R=5$ ms), PC ($\tau_T/\tau_\mw=3$,$\tau_R=8$ ms) and DC ($\tau_T/\tau_\mw=10$,$\tau_R=40$ ms) phases from the experiments. Scale bar is  1mm. Velocity vectors are superimposed to the particles. Particle area fraction and pulse amplitude are constant in all experiments and are equal to $\phi\approx 0.15$ and E$=2.08$ MV/m, respectively.
(c) Angular-averaged normalized two-point correlation vs. radial distance $r$ normalized by particle diameter $d$; top inset: in 2D, showing the spatial periodicity of rotating cluster; bottom inset: characteristic length of two-point correlation functions defined as the point where $S_2$ crosses the horizontal line. (d) Angular-averaged velocity auto-correlation vs. normalized radial distance; top inset: in 2D, showing the periodicity of velocity correlation of rotating and polar clusters in space. Some degree of anisotropy in the velocity auto-correlation of polar cluster is due to insufficient number of clusters in the field of view; bottom inset: characteristic length of velocity auto-correlation functions, defined as the point where $C_{vv}$ crosses the zero line. (e) polar order parameter vs. cluster size of $n$ particle; inset: global order parameter, showing lack of any global ordering in all phases; shaded areas indicate polar ordering higher than $75\%$ and lower than $15\%$.		
		 }}
	\label{fig3}
\end{figure*}

{\it{Collective dynamics:}} The Quincke random walkers exhibit  rich collective dynamics, summarized  in Fig. \ref{fig3} for the case of a simple walk.  
At a given particle density, depending upon the run time $\tau_R$ and the degree of depolarization $\tau_T/\tau_\mw$ (``memory"), the Quincke colloids self-organize into different dynamical phases at (statistically) steady-state with distinct statistical properties (Fig. \ref{fig3}c-e). The classical, run-only Quincke rollers  \cite{Bricard13} correspond to $ \tau_T/\tau_\mw=0$.
The  phases are differentiated by examining the spatial two-point correlation function $S_2$,
velocity auto-correlation function $C_{vv}$,
and the polar order parameter $\Phi_O$ (see Appendix for definitions).
The spatial correlations and order parameters are computed from equal-time averages over the frames corresponding to running periods.   The flow field analysis in the population of particles was performed using an open source Matlab code PIVLab  \cite{Thielicke14}.

\begin{figure*}[t]
	\includegraphics[width=\linewidth]{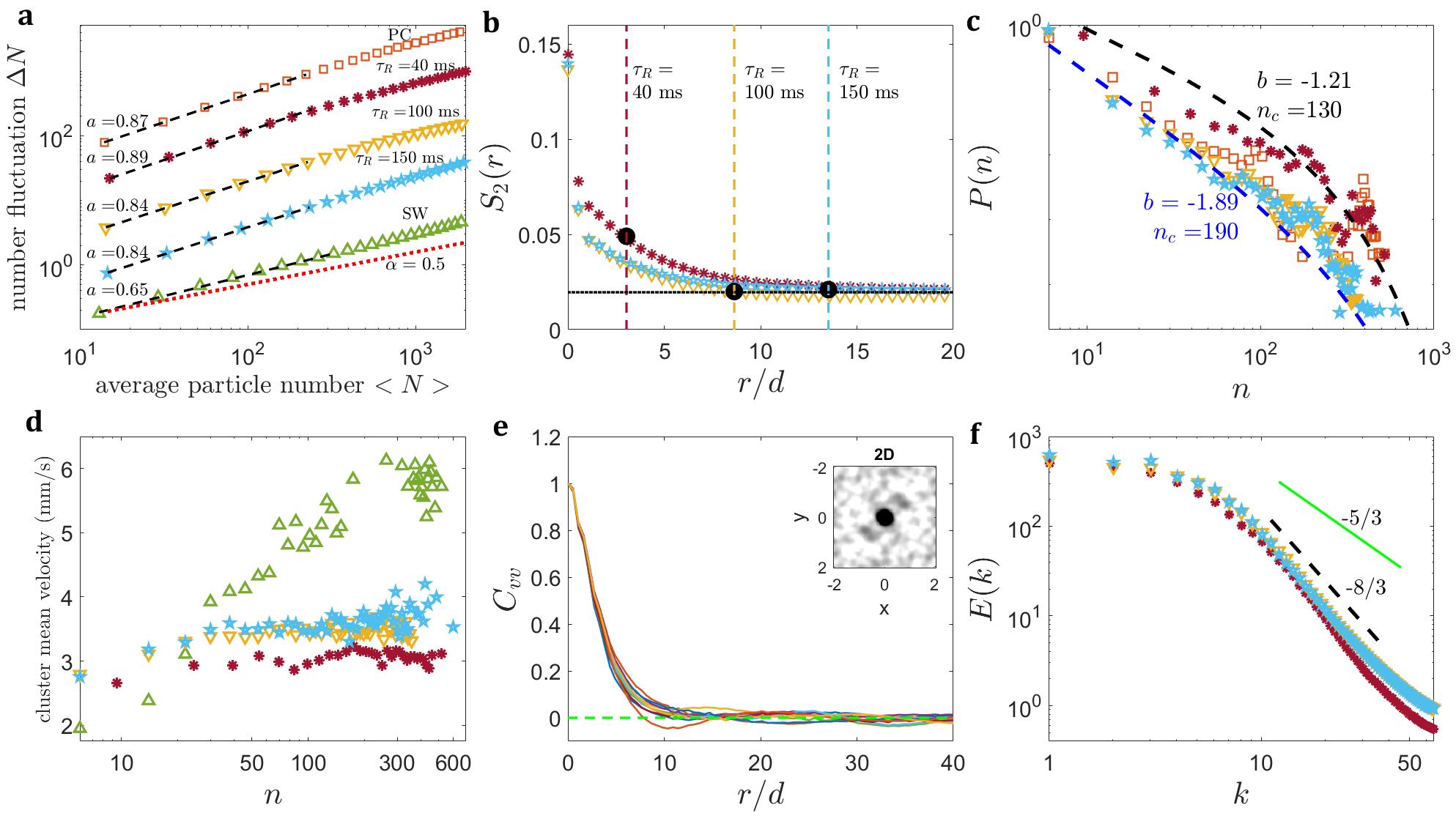}
	\caption{\footnotesize{(a) Number fluctuations $\Delta N$ as a function of average number of particles $\langle N\rangle$. Normal fluctuation is shown by dotted line. Plots shifted vertically
	for better visualization. (b) Spatial two-point correlation of disordered clusters. Vertical lines depict the corresponding kinematic length scales. $\ast$: \ ($\tau_T/\tau_\mw=10$,$\tau_R=40$ ms), $\triangledown$: ($\tau_T/\tau_\mw=10$,$\tau_R=100$ ms), $\star$: ($\tau_T/\tau_\mw=10$,$\tau_R=150$ ms). 
(c) Probability distribution of clusters with {\itshape n} number of particles (d) cluster mean velocity averaged over all the particles in a cluster as a function of cluster size of {\itshape n}. 
(e) Velocity auto-correlation of a disordered cluster with $\tau_R=100$ ms evaluated at different run steps. The anti-correlation around $r/d\approx 10$ shows the approximate size of the highly transient vortical structures formed in disordered clustering. Inset: All run-time averaged velocity auto-correlation  in 2D.  (f) Angular-averaged energy spectrum of disordered clusters formed by Quincke random walkers with different run times $\tau_R$= 40, 100 and 150 ms. \\	
Symbols denote different phases: $\square$: polar cluster (PC) ($\tau_T/\tau_\mw=3$,$\tau_R=8$ ms), $\ast$: disordered cluster (DC) \ ($\tau_T/\tau_\mw=10$,$\tau_R=40$ ms), $\triangledown$: disordered cluster ($\tau_T/\tau_\mw=10$,$\tau_R=100$ ms), $\star$: disordered cluster ($\tau_T/\tau_\mw=10$,$\tau_R=150$ ms), $\bigtriangleup$: swarm (SW) ($\tau_T/\tau_\mw=1$,$\tau_R=4$ ms)
}}
	\label{fig4}
\end{figure*}

If the colloid run directions are correlated due to significant memory effect  ( $\tau_T/\tau_\mw<2$) and  run times are short  $\tau_R\sim\tau_T$, particles form swarms  similar to those observed in \cite{Bricard13,Bricard15}. The fast-decay of $S_2$ in Fig. \ref{fig3}c and the corresponding characteristic length scale $L_{S_2}$ of a few particle diameter $d$, ($L_{S_2}$ defined as the length where $S_2$ crosses the horizontal line corresponding to the  $r \rightarrow \infty$ limit), indicate lack of connectiveness or large scale clustering of the particles. However, swarms show long-range velocity correlations and high polar ordering, see  Fig. \ref{fig3}d and e. 

Increasing $\tau_R$, while keeping  $\tau_T/\tau_\mw$ below 2,  leads to the emergence of stable rotating cluster islands, with (periodic) long-range spatial pair and velocity correlations and very weak polar ordering (Fig. \ref{fig3}). It has been argued that pair-aligning interactions are responsible for local grouping of Quincke rollers  in swarming state \cite{Bricard13,Bricard15,Lu18}, however,  recent particle-based simulations \cite{Grossmann15} suggest that the formation of rotating clusters can be attributed to the enhancement of a competing anti-aligning interaction. 

As the memory effect fades,  $\tau_T/\tau_\mw> 2$, the swarm and rotating clusters disappear. 
At small to intermediate values of $\tau_R$, particles form stable large polar clusters, which cruise over the whole domain without significant exchange of particles. The main characteristic feature of these clusters is the long-range velocity and orientational order across giant moving clusters, as shown in Figs. \ref{fig3}(c-e). 

As both $\tau_T$ and $\tau_R$ increase, the giant polar clusters break up by exchanging outermost particles, which start performing independent random-walks. This results in a more continuous spectrum of cluster size distribution at steady-state, with large clusters being orientationally decorrelated, as shown in Fig. \ref{fig3}e. 
The resulting disordered clusters are highly dynamic: they continuously evolve, deform and break by exchanging particles. 


Clustering is characterized by 
anomalous density fluctuations (Fig. \ref{fig4}a).  The particle number fluctuations $\Delta N$
scales with the average particle numbers $\langle N\rangle$ (in windows of different linear size)
as $\Delta N \sim \langle N\rangle^a$, with exponent $a$ which is larger than the one for fluctuations observed for systems in thermal equilibrium $a=0.5$.
Compared to the disordered and polar cluster phase, the swarming phase lacks long-range clustering, which  results in a more uniform distribution of cells and thus a smaller power-law exponent $a$.
The density fluctuations  are sensitive to the run time. 
Fig. \ref{fig4}a shows that in the disordered cluster phase as $\tau_R$ increases from $40$ms to $100$ms and $150$ms, the power-law exponent $a$ decreases from $0.89$ to $0.84$. The origin can be traced to the relative magnitude of the Quincke roller average run length and the cluster size; if the run is shorter than the aggregate length, the particle remains trapped in the cluster.  Fig.\ref{fig4}b shows that  the run length,  estimated by the kinematic length scale of the random walkers  $L_k=V \tau_R$, 
falls in regions with considerable degree of spatial correlation for the  short run time $\tau_R=40$ms. However, for  $\tau_R=100$ ms and $150$ ms, $L_k$ intersects the corresponding $S_2$ curves at points where spatial correlation almost vanishes. The longer kinematic length scale makes it possible for the outermost particles to leave their original cluster and diffuse into an already existing one or form a new cluster with other isolated random walkers. At even shorter run times, the kinematic length scale becomes much smaller than the pair-correlation length scale, which results in either very low mobility or even static clusters ((upper left part of the phase diagram in Fig. \ref{fig3}a))

 Intriguingly, the 
observed anomalous scaling for the number density fluctuations  in the polar and disordered cases are comparable with those obtained in moving clusters of gliding {\it{M. xanthus}} mutant \cite{Peruani12} and swimming {\it{B. subtilis}} \cite{Zhang10},  thereby suggesting that the disordered clusters behave similarly to ones observed in  bacterial systems. 
Indeed, the exponent of the inverse power-law scaling of the cluster size probability distribution (Fig. \ref{fig4}c), $P(n)\sim n^{b}e^{-n/n_c}$, 
agrees well with dynamic clustering in bacterial suspension and discrete particle simulations \cite{Zhang10,Pohl14}.
Furthermore, as illustrated in Fig. \ref{fig4}d, cluster mean velocity $V_m$ increases with the size of the cluster and plateaus beyond certain cluster sizes, similar to \cite{Zhang10}. The figure also shows that the high orientational ordering in polar clusters significantly enhances the mean velocity, compared with the disordered clusters.  
The angular-averaged velocity auto-correlation of disordered clusters in Fig. \ref{fig4}e shows anti-correlation around $r/d\approx 10-20$, which is a signature for the formation of vortical structures, similar to those observed in different bacterial systems \cite{Dombrowski04,Cisneros07,Zhang09,Cisneros11,Wensink12b,Dunkel13}.
The corresponding energy spectrum calculated from the velocity field of the particles shows scaling of $-8/3$ (see also Fig. \ref{fig4}f), which is in agreement with mesoscopic turbulence in bacterial suspension \cite{Wensink12b}, discrete particle simulations \cite{Grossmann14}, and also in numerical simulations for suspension of pushers in a Newtonian fluid \cite{Li16}. 

The quantitative similarity of the cluster and flow statistics of bacterial and Quincke walker clusters may originate from a unique feature of the Quincke random walkers: when the field is on, they all run and when the field is turned off, they all stop.
This  de facto synchronization of the runs and turns mimics physical locking and intertwining of flagella in dense clusters of bacterial systems and thus may play a  role in the observed coordinated motion\cite{Copeland09,Zhang10}. 

The creation of the Quincke random walker enables the experimental study of active fluids emulating bacterial suspensions under well defined and controllable conditions, e.g., particle density, speed (i.e., activity) and walk type. 
In this work we only focused  on the effects of the simple walk and its characteristics (run and turn times) on the collective dynamics at moderate particle density.
 Exploration of the complete phase space will likely discover more complex
 collective states, for example, 
preliminary results show vortex arrays similar to ones observed in swimming sperm \cite{Riedel05}
and in particle-based simulations  \cite{Grossmann14,Grossmann15}. The Quincke random walker can also be programmed  with alternating or time-varying speed \cite{Theves13,Babel14} and locomotions with distributed waiting times featuring anomalous subdiffusion. 
Another fundamental problem that can be investigated with  the Quincke random walker is the  navigation of biological microswimmers in 
the complex heterogeneous environment  \cite{Morin17, Morin17b}.
This would provide an insight on how different search strategies, such as L\'{e}vy walk, are affected by the presence of obstacles and which intermittent motility pattern yields the optimal search strategy \cite{Zaburdaev15,Volpe17}.  Our approach can also be used to randomize the motion of other active particles powered by the Quincke effect such as the recently proposed helical propeller \cite{Das:2019} and use this microswimmer to explore self-organization in three-dimensional suspensions.We envision the Quincke random walker as a new paradigm for active locomotion at the microscale and a testbed for the abundant theoretical models of the collective dynamics of active matter.

This research has been supported  by NSF awards CBET-1704996 and  CMMI-1740011.

\appendix

\section{Quincke Effect}

The spontaneous spinning of a rigid sphere in a uniform DC electric field $\mathbf{E}$ has been known for over a century \cite{Quincke:1896, Melcher-Taylor:1969, Lemaire:2002}. Yet this phenomenon has been  largely overlooked until its recent application to  power ``active" particles, in particular  the Quincke rollers  \cite{Bricard13,Bricard15, Snezhko:2016, Lavrentovich:2016, Gerardo:2019}. 

\begin{figure}[h]
	\centering
	\includegraphics[width=3in]{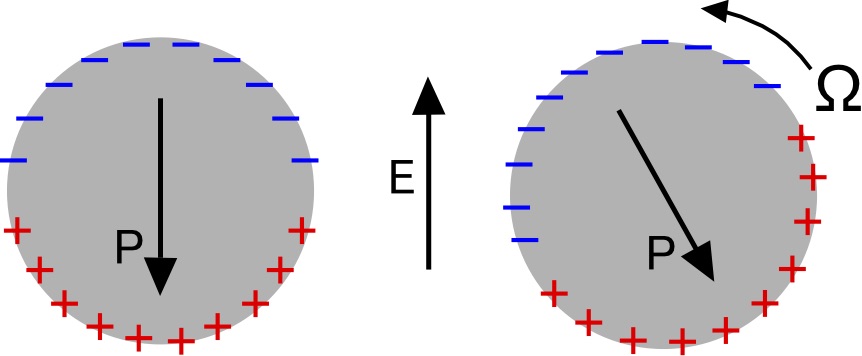}
	\label{Quincke_schematic}
	\caption{Induced free charge distribution  for a sphere with $R/S<1$.  Above a critical field strength $E>E_Q$ steady rotation around an axis perpendicular to the electric field  is induced by the misaligned induced dipole  of the particle (right). The rotation can be either clock- or counter-clockwise. }
	\label{Quinckefigure}
\end{figure}

The phenomenon arises from particle polarization in an applied electric field due to the accumulation of free charges at the particle interface  \cite{Melcher-Taylor:1969}.
The induced dipole due to these free charges lags the application of the uniform DC electric field as
\begin{equation}
\label{Peff}
\begin{split}
\bP=& \left(\chi_0-\chi_\infty \right) \bE \left[1-\exp\left(-t/t_\mw\right)\right]\,, 
\end{split}
\end{equation}
where $\chi_0,\, \chi_\infty$ are the low- and high-frequency susceptibilities of the particle, and $t_{\mw}$ is the Maxwell-Wagner polarization time.
For a   sphere  with diameter $d$ 
\begin{equation}
\label{Peff2}
\begin{split}
\chi_0-\chi_\infty= \frac{ \pi}{2} \eps_\out d^3  \frac{3(\Rr-\Sr)}{(\Rr+2)(\Sr+2)}\,,\quad t_{\mw}=\frac{\eps_{\ins}+2\eps_{\out}}{\sigm_{\ins}+2\sigm_{\out}}\,.
\end{split}
\end{equation}
 $\Rr$ and $\Sr$ characterize the mismatch of  electrical conductivity $\sigm$,  and permittivity $\eps$, between the sphere and  the suspending fluid:
\begin{equation}
\Rr=\frac{\sigm_\ins}{\sigm_\out}\,,\quad \Sr=\frac{\eps_\ins}{\eps_\out}\,,
\label{ParameterRatios}
\end{equation}
where subscripts $``\ins "$ and $``\out "$ denote values for the particle and suspending medium, respectively.

In order for rotation to occur, the induced free-charge dipole of the sphere should be oriented opposite to the direction of the applied field, which occurs when $\Rr/\Sr<1$ (see \refeq{Peff} and \refeq{Peff2}). This configuration is unfavorable and becomes unstable above a critical strength of the electric field. A perturbation  tilts the dipole, leading to an electric torque that drives physical rotation of the sphere around an axis perpendicular to the applied field direction.  For this rotation to be sustained, the rotation period should be comparable to the  Maxwell-Wagner polarization time:    this condition ensures that while the induced surface-charge distribution rotates with the sphere,  the exterior fluid  can recharge the interface by conduction.  The balance between charge convection by rotation and supply by conduction from the bulk results in an oblique dipole orientation  with a steady angle  as shown in Figure \ref{Quinckefigure}.  

The steady rotation rate $\Omega$ of the Quincke rotor is determined from 
conservation of angular momentum of the sphere, where electric and viscous torques balance rotational inertia \cite{Turcu:1987, Jones:1984, Lemaire:2002}:
 \begin{equation}
 \label{AngM}
 \begin{split}
I \frac{d \Omega}{d t}=P_{\perp}E-\zeta \Omega\,,
\end{split}
\end{equation}
Here, $I=\pi \rho_\ins  d^5/60$ is the moment of inertia and $\zeta= \pi \mu_\out d^3$ is the  friction factor of the sphere. The dipole component orthogonal to the field direction, $P_\perp$,  is determined from the coupled evolution equations for the polarization:
\begin{equation}
\label{Pevol}
 \begin{split}
\frac{d P_\perp}{d t}&=-\Omega P_{||}-t_\mw^{-1}P_\perp\,,\\
\frac{d P_{||}}{d t}&=\Omega P_\perp-t_\mw^{-1}\left[P_{||}-\left(\chi_0-\chi_\infty\right)E\right]\,.
\end{split}
\end{equation}
The steady state solutions of \refeq{AngM} and \refeq{Pevol}  are  no rotation ($\Omega=0$ with induced dipole $P_{||} $ given by \refeq{Peff} and $P_\perp=0$), and steady rotation with
 \begin{equation}
\label{quinckeW}
\begin{split}
 \Omega&=\pm \frac{1}{t_{\mw}}\sqrt{\frac{E^2}{E_Q^2}-1}\,,\\
 E_Q^2&=\frac{2\sigm_\out \mu_\out \left(\Rr+2\right)^2}{3\eps_\out ^2 (\Sr-\Rr)}\,, \quad P_\perp=\frac{\zeta \Omega}{E}\,.
 \end{split}
\end{equation}
 where  $\pm$  reflects the two possible directions of rotation, 
 shows that rotation is only possible if the electric field exceeds a critical value given by $E_Q$.

\section{Experiment}
The colloidal rollers are polystyrene spheres ({Phosphorex}, {Inc}.) with diameter $d=40 \mu m$, density $\rho_\ins=1.18$ g/cm$^3$, and dielectric constant $\eps_\ins\sim 3$. The suspending  fluid is hexadecane oil  (Sigma Aldrich), density $\rho_\out$=0.77 g/cm$^3$ and dielectric constant $\eps_\out=2$ ,  containing 0.15 M AOT
(Sigma Aldrich). The conductivity of the solution is  $\sigma_\out=1.2\times10^{-8}$ S/m, measured with a high-precision multimeter (BK Precision).
The Maxwell-Wagner time for this system is $\tau_\mw\sim 2$ ms. 
The experimental setup consists of a  $2 \times 2$ cm$^2$ rectangular chamber made from two Indium-Tin-Oxide (ITO) coated glass slides (Delta Technologies) as electrodes, separated by a  {Teflon} spacer with thickness 120 $\mu$m.
The particle motion and tracking is visualized using an optical microscope (Zeiss) 
mounted on a vibration isolation table ({Kinetic} {Systems}, {Inc}.) and videos were recorded at frame rates higher than 500 frames per second by using a high speed camera (Photron).
The waveform for each walk is  programmed as a Matlab code and  interfaced with a high voltage amplifier (Matsusada) through a function/wave generator (Agilent Technologies).  
Particle tracking and all analyses were performed using custom-written Matlab code. The flow
field analysis in the population of particles was performed using an open source Matlab code PIVLab \cite{Thielicke14}. 

\begin{figure}[h!]
		\includegraphics[width=\columnwidth]{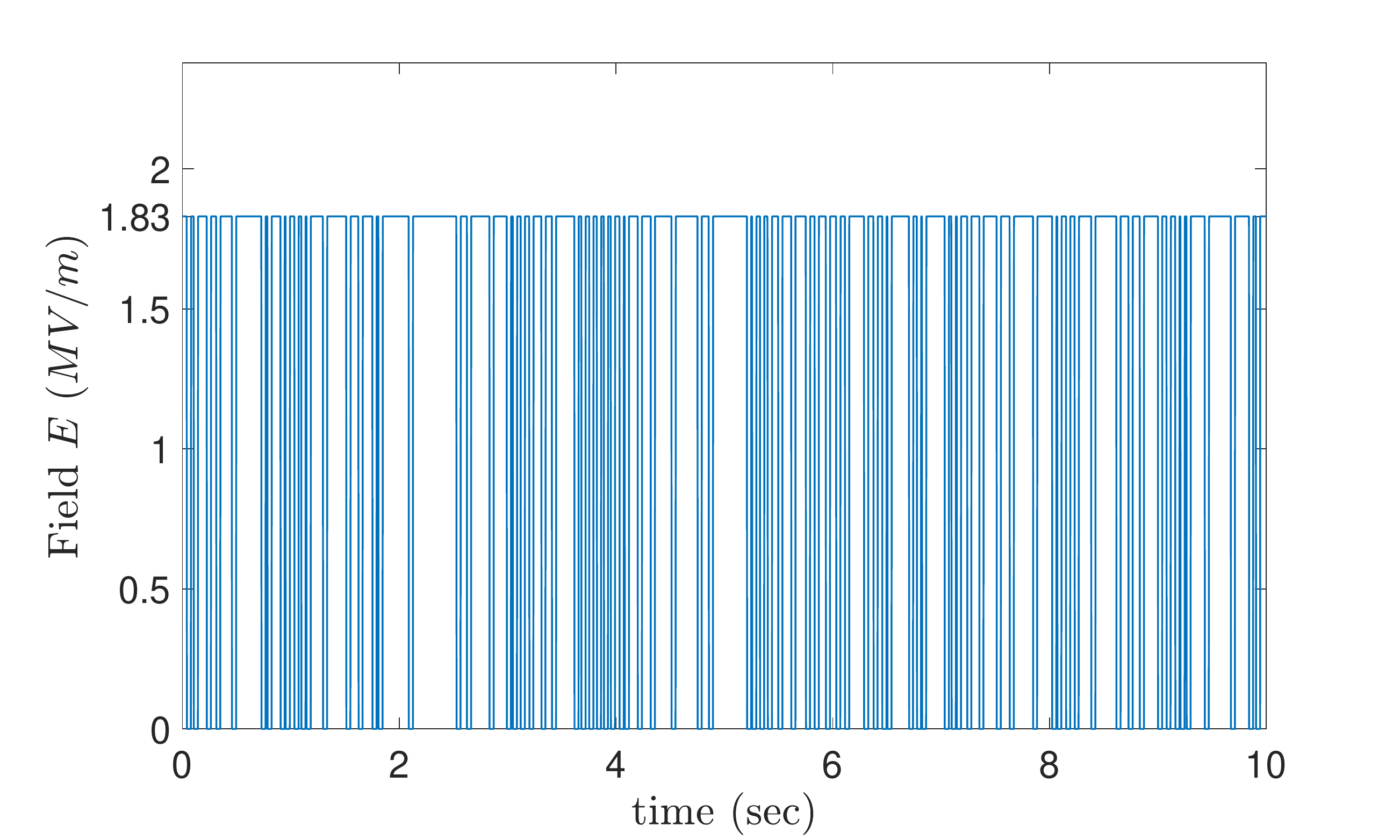}
		\includegraphics[width=\columnwidth]{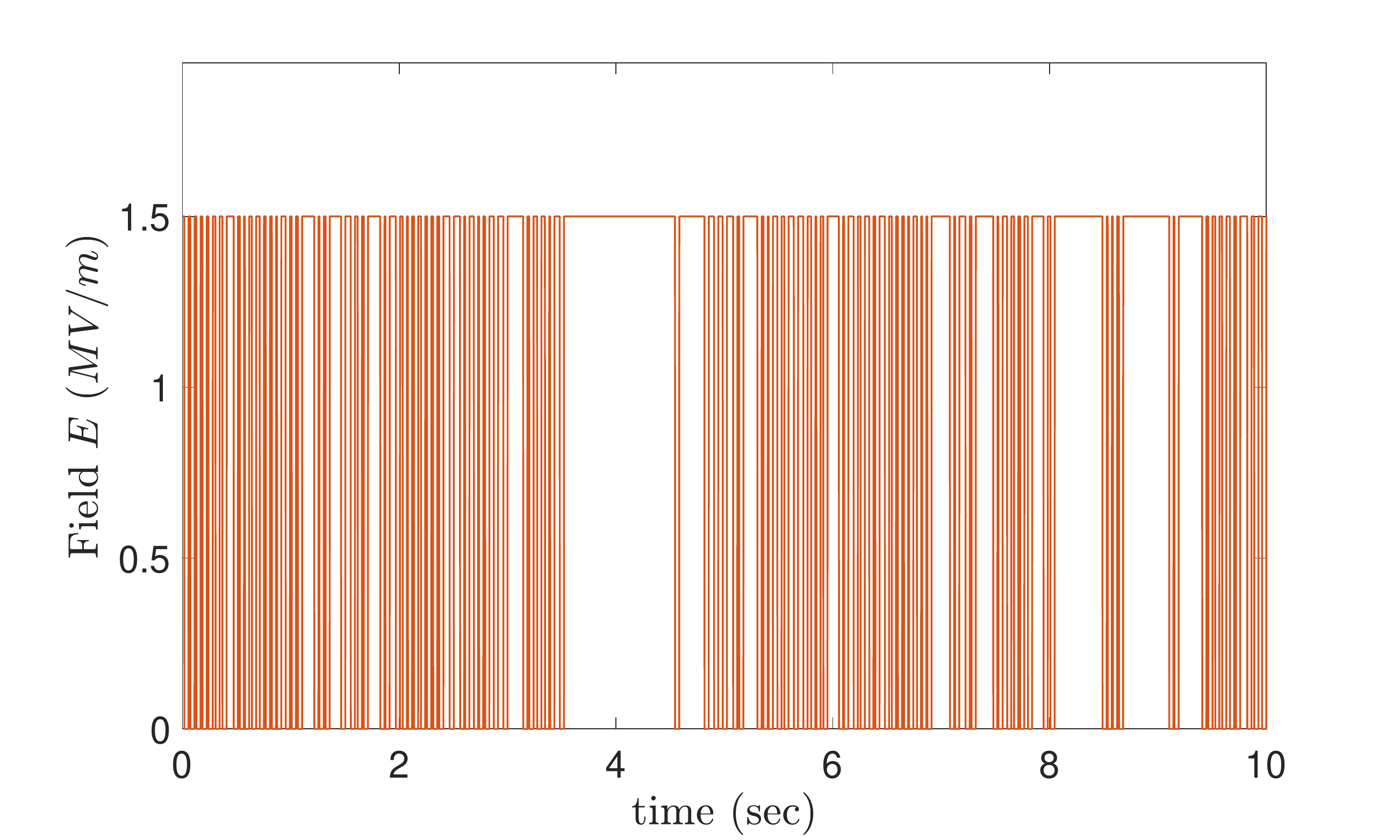}
	\caption{10 second of the generated signal based on the exponential PDF of run-times $\tau_R$ for run-and-tumble motion (a) and power-law PDF for L\'{e}vy-walk (b).}
	\label{fig:signal}
\end{figure}

 \section{Quincke Random Walker {\it vs.} Active Brownian Particle}
In order to compare the present Quincke random walker with active Brownian particles, we estimate the importance of Brownian translational and rotational diffusion by evaluating the P\'{e}clet number, $Pe = V/\sqrt{D_T \times D_R}$ \cite{Bechinger16}, where $D_T=k_B T/(3\pi\mu d)$ and $D_R=k_B T/(\pi\mu d^3)$ are the Brownian translational and rotational diffusivities, and $k_BT$ is the thermal energy. 
For the particle and fluid system used in our experiments, the corresponding R\'{e}clet number 
is $Pe\approx 10^6$, implying that the Brownian diffusion can be ignored on the time scales of our experiments and that the rotational diffusion does not play any significant role in reorienting the Quincke random walker, in contrast to the artificial active Brownian particles.

The choice of particle size and the selected velocity range in this report are set by the limitations in the spatial and temporal resolutions of our experimental setup and a compromise for minimizing the localization error in single particle tracking.
It is interesting to perform experiments where the contribution of Brownian diffusion is superposed on the run and turn events \cite{Thiel12,Detcheverry17}, similar to situations experienced by biological microswimmers. 
This can be achieved for example, through lowering the P\'{e}clet number by using smaller size particles, e.g. $1 \mu m$, and lower velocities in the Quincke random walk.

\section{Theoretical MSD and VACF for individual random walkers}

For any general run time distribution functions $\psi_R(t)$, we first compute the corresponding distributions in the Laplace domain, i.e. $\psi_R(s)$, and use the formulations presented in \cite{Angelani13,Detcheverry17} to find the mean-squared displacement $MSD(t)=\left\langle \left[ {\bf x}(t+t') -{\bf x}(t')  \right] ^2   \right\rangle $, where {\bf x} is the position of the particle and the brackets denote time averaging with respect to the time variable $t'$ for each trajectory. 
In all cases, we assume a constant run velocity $V$ and turning time $\tau_T$.
Also, because the P\'{e}clet number for the Quincke random-walker is sufficiently high in all our experiments, we ignored the contribution from the Brownian diffusion in the theoretical MSD derivations.

For a random walk with a {\it constant run time $\tau_R$}, the corresponding distribution $\psi_R$ will be equal to $\psi_R(t)=\delta(t-\tau_R)$, or equivalently $\psi_R(s)=e^{-s\tau_R}$ in the Laplace domain. 
Then, we find the mean-squared displacement as:
\begin{equation}\label{eq:msd_unf}
\begin{split}
MSD(t) & = \frac{V^2}{3(\tau_R+\tau_T)}\left[ 3\tau_R  t^2-t^3 \right] \quad \textrm{for $t<\tau_R$}\ \\
 & = \frac{V^2}{3(\tau_R+\tau_T)} \left[ \tau_R^2 (3t-\tau_R) \right]  \quad \textrm{for $t>\tau_R$}\
\end{split}
\end{equation}
 
For the {\it run-and-tumble} motion with a mean run time $\overline{\tau}$, the run times are drawn from an exponential distribution $\psi_R(t)=1/\overline{\tau}\ e^{-t/\overline{\tau}}$, or equivalently $\psi_R(s)=1/(1+s \overline{\tau})$ in the Laplace domain. The mean-squared displacement is readily given by \cite{Angelani13,Detcheverry17}:
\begin{equation}\label{eq:msd_exp}
MSD(t) = \frac{2\ V^2\ \overline{\tau}^2}{1+\tau_T/\overline{\tau}} \left[ e^{-t/\overline{\tau}} + t/\overline{\tau} -1   \right].
\end{equation}

For the {\it L\'{e}vy walk}, we extract the run times from a power-law distribution of the form $\psi_R = \gamma \ t_0^\gamma \ t^{-(1+\gamma)} \ H(t-t_0)$, where $H$ is the Heaviside function, $t_0$ is the lower cutoff for the run times, and $1<\gamma<2$ controls the extent of superdiffusive behavior of the walker that emerges at long times. 
The mean run time $\overline{\tau}$ becomes $\overline{\tau}=t_0 \ \gamma/(\gamma-1)$. 
For $1<\gamma<2$, the mean-squared displacement is given by \cite{Angelani13}:

\begin{equation}\label{eq:msd_pl}
\begin{split}
MSD(t) & = \frac{V^2t^2}{(\overline{\tau}+\tau_T)}\left[ \overline{\tau} -t/3 \right] \quad \textrm{for $t<t_0$}\ \\
& = \frac{V^2}{(\overline{\tau}+\tau_T)} \left[\frac{\gamma t_0^3}{3(3-\gamma)}   - \frac{\gamma t_0^2}{2-\gamma}t \right.\\
&  \left. + \frac{2t_0^\gamma}{(\gamma-1)(2-\gamma)(3-\gamma)}t^{3-\gamma}         \right]  \quad \textrm{for $t>t_0$}\
\end{split}
\end{equation}

To find the theoretical temporal velocity auto-correlation function $VACF(t) = \left\langle {\bf v}(t+t') \cdot {\bf v}(t') \right\rangle $, we use  \refeq{eq:msd_unf}-\refeq{eq:msd_pl} along with the relation d$MSD(t)$/d$t$ = $2\times \int VACF\ dt$.
For the run-and-tumble motion with run times drawn from an exponential pdf, temporal VACF becomes:
 
\begin{equation}\label{eq:vac_exp}
VACF(t) = \frac{V^2}{1+\tau_T/\overline{\tau}}\ e^{-t/\overline{\tau}},
\end{equation}

And for the L\'{e}vy walk with the power-law distribution of run times, we obtain the temporal VACF as:

\begin{equation}\label{eq:vac_pl}
\begin{split}
VACF(t) & = \frac{V^2}{(\overline{\tau}+\tau_T)}\left[ \overline{\tau} -t \right] \qquad \textrm{for $t<t_0$}\ \\
& = \frac{V^2}{(\overline{\tau}+\tau_T)} \left[\frac{t_0^\gamma}{(\gamma-1)}t^{1-\gamma}         \right]  \quad \textrm{for $t>t_0$}\
\end{split}
\end{equation}

For calculating the experimental velocity auto-correlation function from the particle trajectory, we use the Wiener-Khinchin autocorrelation theorem \cite{Bracewell65}. The experimental velocity auto-correlation then can be accurately calculated in the Fourier domain by taking the inverse Fourier Transform of $|FT({\bf v}(t))|^2$, i.e. absolute square of the Fourier transform $FT$ of the particle velocity {\bf v}(t).

\section{Cluster identification, Spatial correlation functions and Order parameters}

We define clusters as a group of all particles located at a distance closer than some threshold value, regardless of their orientation. Several trials show that binning based on a threshold distance of $1.4d-1.6d$, where $d$ is the diameter of a single colloid, would result in a correct identification of clusters. 

The standard two-point correlation function $S_2$ provides a robust measure for calculating the probability of finding two points of a microstructure ${\bf x_1}$ and ${\bf x_2}$ both in the same phase, which for our purposes is the particle phase. 
It is defined as \cite{Torquato13}:
\begin{equation}\label{eq:S2}
S_2({\bf x_1},{\bf x_2}) = <I({\bf x_1}) \cdot I({\bf x_2})>,
\end{equation}
where angular brackets denote an ensemble average over all possible pairs in space and $I$ is the indicator function or density phase field of the particle, having a value of 1 if it falls in particle and zero everywhere else. 
In case of a statistically isotropic system, $S_2({\bf x_1},{\bf x_2})$ van be angularly averaged to give $S_2(r)$, where the scalar $r$ is the distance between two points.
The descriptor $S_2(r)$ contains a wealth of information regarding the connectivity of different phases in a microstructure. 
$S_2$ can be easily calculated using the Fourier transform of the binary field of each image. We take the average of $S_2$ from all the frames belonging to the running-period.

In order to extract the spatial (clustering) length scales of different patterns, we merge (bin) all the particles belonging to the same cluster and then compute $S_2$ for the binned binary image. This will provide an average cluster size that we observe in each dynamical pattern. 
The corresponding length scale $L_{S_2}$ is the point where $S_2$ curve starts to plateau. 
According to it's definition in Eq. \ref{eq:S2}, $S_2(r=0)$ is equal to the particle volume fraction $\phi$ and $S_2(r\rightarrow\infty)$ will be equal to the joint probability of satisfying simultaneously $I({\bf x_1})=I({\bf x_2})=1$, which is $\phi^2$ \cite{Torquato13}. We use both $S_2$ and normalized $S_2$, i.e. $S_2/S_2(r=0)$ interchangeably in our quantitative analysis.  

The velocity auto-correlation $C_{vv}$ is calculated from the following relation:
\begin{equation}\label{eq:Cvv}
C_{vv}({\bf x_1},{\bf x_2}) = <{\bf v}(t,{\bf x_1}) \cdot {\bf v}(t,{\bf x_2})>,
\end{equation} 
where angular brackets denote spatial averaging over all possible pairs.
In case of a statistically isotropic system, we use the angular-averaged $C_{vv}(r)$, where $r$ is the distance between two points over the space. We use the normalized velocity auto-correlation in our analysis, i.e.  $C_{vv}(r)/C_{vv}(r=0)$. 
From the velocity auto-correlation $C_{vv}({\bf x_1},{\bf x_2})$, we compute the 2D Fourier transform to get the energy spectrum $E_2(k_x,k_y)$, where $k_x$ and $k_y$ are wave-numbers in $x$ and $y$ directions, respectively. In case of a statistically isotropic field, we average $E(k_x,k_y)$ over different wave-number angles to get $E(k)$.

The order parameter $\Phi_O$ of a cluster containing $n$ particles is calculated from $\Phi_O=|<e^{i\theta}>|$, where $\theta$ is the instantaneous direction of motion of each particle in a cluster and $<.>$ is the average over all the particles in a specific cluster.

\bibliographystyle{unsrtnat}

\end{document}